\documentclass[sigconf,10pt]{acmart}
\usepackage{xcolor, soul}
\sethlcolor{cyan}

\AtBeginDocument{%
  \providecommand\BibTeX{{%
    \normalfont B\kern-0.5em{\scshape i\kern-0.25em b}\kern-0.8em\TeX}}}

\copyrightyear{2020}
\acmYear{2020}
\setcopyright{acmlicensed}
\acmConference[SIGMOD'20]{Proceedings of the 2020 ACM SIGMOD International Conference on Management of Data}{June 14--19, 2020}{Portland, OR, USA}
\acmBooktitle{Proceedings of the 2020 ACM SIGMOD International Conference on Management of Data (SIGMOD'20), June 14--19, 2020, Portland, OR, USA}
\acmPrice{15.00}
\acmDOI{10.1145/3318464.3386135}
\acmISBN{978-1-4503-6735-6/20/06}

\begin{document}
\fancyhead{}

\title{A1: A Distributed In-Memory Graph Database}

\author{Chiranjeeb Buragohain, Knut Magne Risvik,
Paul Brett, Miguel Castro,
Wonhee~Cho, Joshua Cowhig, Nikolas Gloy,
Karthik Kalyanaraman, Richendra~Khanna, John Pao,
Matthew Renzelmann, Alex Shamis, Timothy Tan, Shuheng Zheng}
\authornote{
Chiranjeeb Buragohain and Richendra Khanna are currently at Oracle, Timothy Tan is currently at Amazon}
\affiliation{%
  \institution{Microsoft}
}


\renewcommand{\shortauthors}{Buragohain et al.}

\begin{abstract}
A1 is an in-memory distributed database used by the Bing search engine to support complex queries over structured data.  The key enablers for A1 are availability of cheap DRAM and high speed RDMA (Remote Direct Memory Access) networking in commodity hardware.  A1 uses FaRM \cite{farm-nsdi, farm-sosp} as its underlying storage layer and builds the graph abstraction and query engine on top. The combination of in-memory storage and RDMA access requires rethinking how data is allocated, organized and queried in a large distributed system. A single  A1 cluster can store tens of billions of vertices and edges and support a throughput of 350+  million of vertex reads per second  with end to end query latency in single digit milliseconds.  In this paper we describe the A1 data model, RDMA optimized data structures and query execution.
\end{abstract}

\maketitle

\section{Introduction}\label{sec:introduction}
%

The Bing search engine handles massive amounts of unstructured and
structured data. To efficiently query the structured data, we need a
platform that handles the scale of data as well as the strict
performance and latency requirements.
A common pattern for solving low-latency query serving is to use a
two-tier approach with a durable database for the ground truth 
and a caching layer like memcached in front for read-only serving. The
Facebook TAO datastore\cite{fb-tao} is a sophisticated example of this
architecture using the graph data model. But there are some elements
of the design that introduce problems. First, systems like
memcached expose a primitive key-value API with little query
capability.  Therefore complex query execution logic is pushed into
the client, rather than the database itself.  Second, cache
consistency is hard to achieve and such systems guarantee only
eventual consistency.  Finally, there is no atomicity for updates
which leads to data constraint violations.  For example, in TAO one can
have partial edges between two objects with the forward link existing,
but no backward link.  In Bing, we have a huge set of diverse data
sources that need be stitched together with real-time update
requirements.  Therefore we wanted to move beyond an eventually
consistent cache and into a more capable transactional database system.

In representing structured data, the relational and the graph data
models have equivalent capabilities, though with different ease of 
expression\cite{robinson2013graph}.  Our choice of the graph data model is a natural
match for much of Bing data including core assets like the knowledge
graph\cite{knowledge-graph}.  Therefore we designed A1 to be a general
purpose graph database with full transactional capabilities. 
Transactions in a distributed system frees up application developers from
worrying about  complex problems like atomicity, consistency
and concurrency control, and instead allows them to focus on core business
problems\cite{spanner}.  A1
also exposes a query language which simplify application
development by moving query execution into the database.  Our query
language doesn't attempt to be as comprehensive as SQL and instead
focus on the core capabilities needed by the applications using
A1. The primitives we support are general enough that multiple classes
of applications can start using A1 with little difficulty.  Another
key characteristic of A1 is that it is a latency-optimized
database. Since search engines like Bing have a fixed latency budget
to render pages, all queries issued to the backend by the search
engine come with a corresponding latency budget (typically 100ms). If
a query takes more than 100ms to execute, then the results of that
query will simply be discarded.  That means that the availability of
the system is measured by its latency, not by its error rate  ---if a system's 80th percentile latency is 100ms,
the system's effective availability is only 80\%.  Therefore having
tight control over tail latency is a key requirement.

Cheap DRAM and fast networks with Remote Direct Memory Access (RDMA)
are the two major hardware trends that enable A1. We are deploying
machines with hundreds of gigabytes of DRAM, so a set of racks can
hold more than 200 TB of memory. This is sufficient for most
applications to keep their data in memory and to avoid accesses to
secondary storage. Until recently, RDMA has been mostly in the province of
exotic high performance computing networks, but now it has become a
commodity technology easily deployed in cloud data centers. The RoCE
(RDMA over Converged Ethernet) networks we use offer a round-trip
latency less than 5 microseconds, bandwidths of 40Gb/s and message
rates approaching 100 million messages per second. Note that running
RDMA in data centers is still not an easy endeavor and we will have
more details on this later. The combination of in-memory storage and
RDMA allows A1 to achieve single digit millisecond latencies for
queries that access thousands of objects across multiple
machines. 

This paper makes three contributions. First, we describe the design and implement of A1 on top of the FaRM distributed memory storage system
(Sections \ref{sec:system-architecture}, \ref{sec:data-structures}).
Next in section \ref{sec:replication}, we show how A1 is integrated into a more complex
system in Bing with replication and disaster recovery. Finally, we evaluate the applications
built on top of A1 and their performance
(Section \ref{sec:applications} and \ref{sec:performance}).  A key
part of our journey in building A1 has been the evolution of a
research prototype like FaRM into a production system.  The learnings
on this path will be described throughout the paper.

\section{System Architecture}\label{sec:system-architecture}
\begin{figure}[htbp]
    \centering
    \includegraphics[width=0.8\columnwidth]{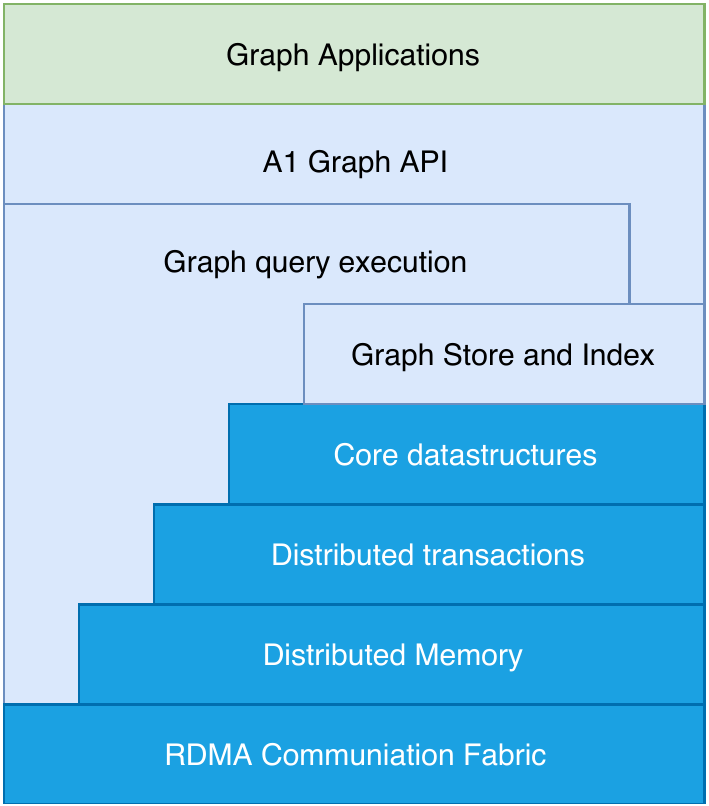}
    \caption{Layers of the A1 architecture}
    \label{fig:a1layers}
\end{figure}

A1 has a typical layered architecture with networking at the bottom and  query processing at top, as depicted in Figure \ref{fig:a1layers}. 

The four lowest layers of the stack together form a distributed storage platform called FaRM \cite{farm-nsdi,farm-sosp,farm-sigmod}. FaRM provides transactional storage and generic indexing structures, while the rest of A1 provides graph data structures and a specialized graph query engine.  The bottom RDMA communication layer provides primitives like one-sided RDMA read/write and a fast RPC implementation. The distributed memory layer exposes a disaggregated memory model where the API enables one to allocate/read/write objects across a cluster of machines. These objects are replicated so that single machine or rack failure never leads to any data loss.  Given a handle to an object, a single one-sided RDMA read is sufficient to retrieve the object. The transaction engine provides atomicity, failure recovery, and concurrency control. FaRM also exposes basic data structures like B-trees. We use these layers to build a database that exposes the graph data model. The query processing works directly on the graph storage, but it also leverages aspects of the distributed memory platform and communication to scale out and coordinate execution of queries.

Before we get into the details of the graph storage, it is worthwhile to understand the lower FaRM layer in a bit of more detail.  We refer the reader to the existing literature \cite{farm-nsdi, farm-sosp, farm-sigmod} on the implementation of FaRM and instead focus here on building applications on top of FaRM.

\subsection{A Brief Tour of FaRM}
FaRM is a transactional distributed in-memory storage system. It is worthwhile unpacking these 
adjectives.  FaRM uses a set of machines in a datacenter and exposes there combined memory as
a single flat storage space. The storage API exposed by FaRM is very simple: every storage 
\emph{object} in FaRM is an unstructured chunk of contiguous memory.  Objects are uniquely
identified by a 64bit \emph{address} or \emph{pointer} and can range from size 64 byte to 1MB. 
All object manipulations happen in the context of a \emph{transaction}. For durability FaRM replicates all data 3-ways.

FaRM uses RDMA capable NICs (Network Interface Cards), which are becoming commodity in modern data centers for cross-machine communication. 
RDMA enables the ability to read/write the contents of a remote machine's memory with low latency 
($<5 \mu s$ within a rack)  and high throughput. RDMA achieves this low latency in three ways: first in 
the local machine, the RDMA library bypasses the OS kernel and talks directly to the NIC. 
Second, in the remote machine, the memory is accessed by the remote NIC directly without involving the CPU. 
This is known as a \emph{one-way read/write}. Finally, TCP features like reliability and congestion control are all implemented
within the NIC and the network switches which reduce the load on CPU further. 
Of course, taking an ordinary storage system and simply porting its network layer to RDMA doesn't always 
result in high performance. FaRM optimizes its whole stack including replication, transaction protocol 
and data structures to leverage RDMA at every layer to provide a  high performance storage system.

A FaRM cluster is a set of machines each running a FaRM process. One machine is designated as a 
\emph{Configuration Manager(CM)} whose purpose is to keep track of machine membership in the cluster and data placement. 
The memory of each machine is split into 2GB chunks known as \emph{regions}.  Objects are allocated within a 
region and every region is replicated 3-ways in-memory for fault tolerance.  Replication is done using a primary-backup 
mechanism and all reads/writes are served from the primary only which ensures consistency of all operations. The the
64-bit address of an object essentially consists of two 32 bit numbers: the \emph{region id} which uniquely 
identifies the region and the offset within the region where that object is located. The CM is responsible for 
determining which machines are part of the cluster (i.e. membership) and region metadata: allocation of regions 
to machines.  Given a FaRM address, the CM metadata can be used to find the machine which hosts the primary copy of 
the region and then we can use RDMA to directly read the contents of the object by using the offset. All reads and 
writes happen in the context of a transaction.  The transaction protocol is a variant of two phase commit with multiple 
optimizations for RDMA. For example, reads are always done using one-sided RDMA reads which bypass the CPU.  Similarly 
data replication happens using one-sided writes, again bypassing the CPU.  In production deployments, we deploy FaRM machines across at least three fault domains.  A single fault domain consists of a set of machines which share 
a common critical component like a network switch or power supply.  Therefore all machines in a single fault domain may 
become inaccessible in case of a hardware failure.  By replicating data across three fault domains, we ensure that 
no single component failure can lead to loss of more than one copy of the data.

\begin{figure}[!htbp]
\begin{verbatim}
std::unique_ptr<Transaction> CreateTransaction();
ObjBuf* Transaction::Alloc(size_t size, Hint hint);
Addr ObjBuf::GetAddr();
ObjBuf* Transaction::Read(Addr addr, size_t size);
ObjBuf* Transaction::OpenForWrite(ObjBuf *buf);
void Transaction::Free(ObjBuf *buf);
Status Transaction::Commit();
Status Transaction::Abort();
\end{verbatim}
    \caption{FaRM API}
    \label{fig:farm_api}
\end{figure}

The FaRM API(Figure \ref{fig:farm_api}) exposes a set of basic operations on objects: allocation, 
reading, writing  and freeing them.  The \texttt{ObjBuf} object referred to in the API is the wrapper around the FaRM object. Reads return an \texttt{ObjBuf} 
object which holds the data for the object read.  The operations must be executed in the context
of a transaction which provides programmers with atomicity and concurrency control. FaRM transactions provide \emph{strict serializability} as the default isolation level using multi-version concurrency control\cite{farm-sigmod}. Every  transaction has a timestamp associated with it and this timestamp ensures a global order among all the transactions in the system.

 Note that the \texttt{Alloc} API takes a \texttt{Hint} parameter.  The hint is used to determine where to allocate the object: by default we allocate the object in the local machine where the API is invoked.  The more useful option is to pass the address of an existing object in the hint ---in that case we attempt to allocate the object in the same region in which the existing object exists. Since the region is our unit of replication, if two objects are allocated in the same region, they are guaranteed to be on the same machine in spite of machine failures.  The hint is advisory only: in case the region doesn't have enough space, the allocator will find another place to allocate it.

Here is an example of atomically incrementing a 64 bit counter which is stored in FaRM:
\begin{figure}[!h]
\begin{verbatim}
Status status = COMMITTED;
do {
  std::unique_ptr<Transaction> tx = CreateTransaction();
  ObjBuf *rbuf = tx->Read(address, sizeof(uint64_t));
  uint64_t value = *(uint64_t*) buf->data();
  value++;
  Objbuf *wbuf = tx->OpenforWrite(rbuf);
  memcpy(wbuf->data(), &value, sizeof(value));
  status = tx->Commit();
} while (status != COMMITTED);
\end{verbatim}
    \caption{Atomic increment of a counter using FaRM API}
    \label{fig:farm_atomic_increment}
\end{figure}

In this example(Figure \ref{fig:farm_atomic_increment}), we read a FaRM object identified by \texttt{address} 
and extract the value stored in it.  The \texttt{ObjBuf} object is a local immutable copy of the object.  
To modify it, we need to create a writable copy which we do with the \texttt{OpenForWrite} API.  Once all the 
objects have been modified, we commit the transaction which atomically makes the update. The reason we have 
the loop here is that FaRM transactions run under optimistic concurrency control and hence may abort under 
conflict and it is necessary to retry them.  Note that in this model all transaction writes are buffered 
locally.  The \texttt{OpenForWrite} operation doesn't cause any remote operations, it merely creates a 
modified buffer and stores the updates locally.  The \texttt{Commit} operations pushes writes to remote
machines, performs concurrency checks and finally commits the data.

\subsection{Design Principles}

Applications like A1 are integrated with FaRM using what we call
the \emph{coprocessor} model.  In this model, A1 is
compiled into the same executable as the FaRM code and is part of the
same address space.  So calling the FaRM API (Figure  \ref{fig:farm_api})
is as simple as making a regular function call. As part of being a
coprocessor, the application needs to integrate with FaRM's threading
model, which we will talk about later.
The availability of transactions in the FaRM layer proved to be a
great engineering productivity boost in building A1.  When
writing A1, the following principles guided our development:

\begin{itemize}
    \item \emph{Pointer linked data structures}: The standard way to build data structures in FaRM is to use FaRM objects connected by 
    pointers, e.g. linked lists, BTrees,  graphs etc. Since dereferencing a pointer generally require an RDMA read 
    (unless the object is hosted on the local machine), we optimize the layout and placement of data structures to reduce the number of 
    pointer dereferences.  For example, we prefer arrays to store list-oriented data instead of traversing linked lists.
    BTrees with high branching ratio works well for search structures, and we use the tuple $\langle$\emph{address,size}$\rangle$ as the
    pointer which indicates both the address and size of the RDMA read to access the data stored in the object.
    
    \item \emph{In-Memory Storage} : Since the cost of memory is high compared to SSD, we need to be frugal with storage.  
    Typically A1 is used as a fast queryable store with non-queryable attributes stored in cheaper storage systems.  For example, 
    if we are storing the profile of an actor in A1, the photo of the actor will not be stored in A1 itself.
    
    \item \emph{Locality}: RDMA reduces latency, but there is still a 20x-100x difference between accessing local 
    memory vs. remote memory. Therefore, at object creation time, we attempt to co-locate data that is likely to be accessed together in the 
    same machine. Similarly, at query time, we ship query to data to reduce the number of remote reads.  When we reallocate any 
    object, we keep its locality intact by passing the old object's address into the \texttt{Alloc} call.
    
    \item \emph{Concurrency}: Since FaRM transactions run under optimistic concurrency control, it is critical to avoid single points of contention. 
    For read-only queries, we use snapshot isolation to ensure that updates to data does not delay or block read-only operations. 
    When we run a distributed query, all objects across the cluster are read as of a single consistent snapshot 
    version and those versions are not garbage collected until the query runs to completion.
    
    \item \emph{Cooperative Multithreading}: Recall that we compile the application with FaRM itself into a single binary.  Inside the
    FaRM process, coprocessors must run using cooperative  multithreading to share compute resources.  At startup, we allocate a fixed number of threads and
    affinitizes them to the cores.  FaRM code and the application code (i.e. the coprocessor) share these threads. Coprocessors use a
    fixed number of fibers per thread to achieve cooperative multithreading.  All FaRM API calls which touch remote objects are
    asynchronous, but the use of fibers hide the asynchrony and gives the application writer the illusion of writing synchronous code.
\end{itemize}
\begin{figure}[htbp]
    \centering
    \includegraphics[width=0.8\columnwidth]{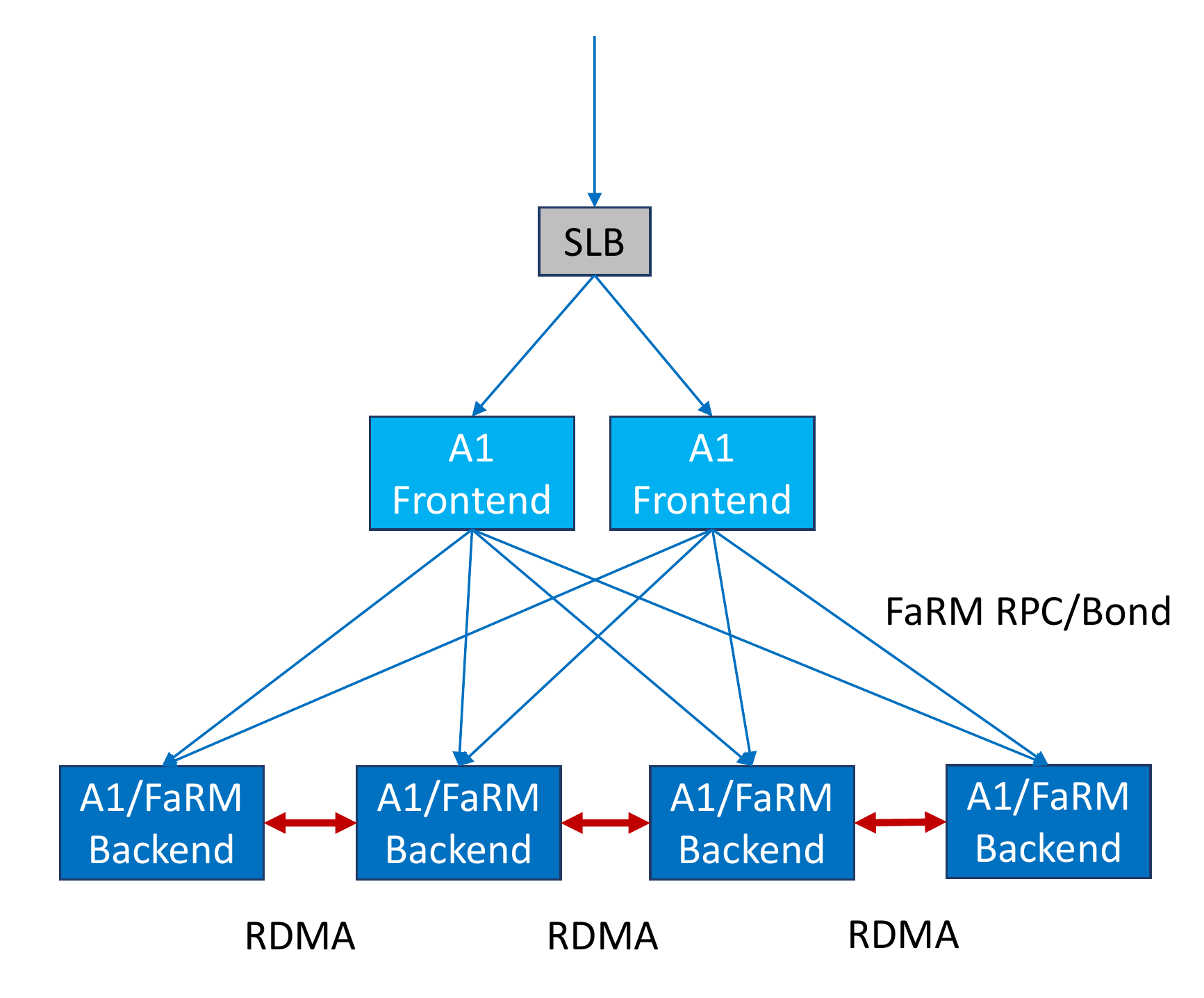}
    \caption{A1 cluster deployment}
    \label{fig:a1machines}
\end{figure}
Figure \ref{fig:a1machines} shows the full physical deployment of a single A1 cluster. Clients access the A1 cluster by making RPC calls to the
A1 API. The RPC calls are routed by a software load balancer (SLB) to a set of frontend machines.  The frontend machines are stateless and 
mostly perform simple routing and throttling functions. The RPC request is forwarded by the frontends to the backend machines where it is 
processed.  The backend machines makes up the FARM cluster, and each machine runs a combined binary of FaRM and A1. All query execution and data
processing happens on the backend machines, utilizing RDMA communication. Communication between client and the cluster uses the traditional TCP stack which has higher latency.
However, our target workloads are complex queries with many reads and writes, so the latency between the client and the backend is typically immaterial
to the total execution time.

\section{Data Structures and Query Engine}\label{sec:data-structures}
\begin{figure}[H]
    \centering
    \includegraphics[width=0.8\linewidth]{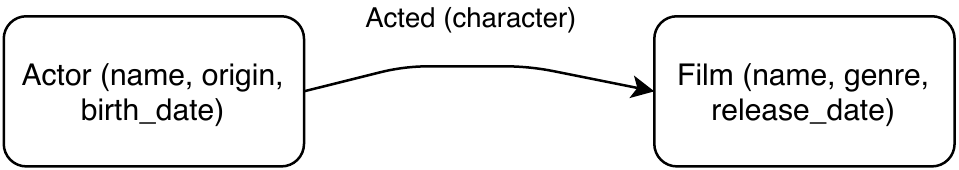}
    \caption{Simple graph example}
    \label{fig:actor-movie-example}
\end{figure}

Using a graph to model data is nothing new --entity-relationship databases have been in use for a while and have enjoyed renewed popularity recently.  
A1 adopts the property graph model: a graph consists of a set of \emph{vertices} and \emph{directed edges} connecting the vertices. The vertices 
and edges are typed and can have attributes (also known as properties) associated with them. The type for the vertex/edge defines the schema of the 
associated attributes.  In contrast to typical property graph models such as Tinkerpop\cite{tinkerpop} or Neo4J\cite{neo4j},
 we choose to enforce schema on attributes to improve data integrity and performance. 
  An example will clarify the model. Consider the relationship between a film and an actor as shown in Figure \ref{fig:actor-movie-example}.

We introduce two types of vertices: \emph{Film} and \emph{Actor}.  The Actor attributes are name, origin and DOB; while the film attributes are name, release date and genre. 
The attributes in the schema are comparable to column definitions in traditional relational databases. Microsoft Bond\cite{bond} is a language for managing schematized data, similar
to Protocol Buffers\cite{protobuf}. In Bond, our schema looks like as follows: 
\begin{verbatim}
struct Actor {                    struct Film {
    0: string name;                 0: string name;
    1: string origin;               1: string genre;
    2: date birth_date;}            2: date release_date;}
\end{verbatim}

Next we introduce the edge type {\em Acted}. that stores data like the name of the character played by the Actor. So the edge data schema will look like
\begin{verbatim}
struct Acted {
    0: string character;}
\end{verbatim}
By using Bond, A1 inherits the Bond type system with primitive types like integers, floats, string, boolean and binary blobs. Since Bond 
allows composite types (arrays and maps) and nesting of structs, A1 can support a richer type system than typical relational 
database. 

A1 organizes customer data in a hierarchy: the top level of the hierarchy is a \emph{tenant} and it is the default isolation container.  Two tenants 
can't see each others data. A tenant may have one or more \emph{graphs} and every graph contain a set of \emph{types}. A graph contains a set of 
vertices and edges and every vertex/edge must belong to one of the types defined within that graph.  The analogy between relational data model and
the A1 model is presented in table \ref{tab:relational_analogy}.

\begin{table}[h]
    \centering
    \begin{tabular}{|l|l|l|l|l|}
    \hline
    \hline
        A1 &  Graph & Type & Vertex/Edge & Attribute \\
        \hline
        Relational & Database & Table & Row & Column \\
    \hline         
    \end{tabular}
    
    \caption{Analogy between relational database entities and A1 entities.}
    \label{tab:relational_analogy}
\end{table}

When declaring a vertex type, the user must also define one of the attributes as a \emph{primary key}, which must be unique and non-null.  Every type by 
default comes with a sorted \emph{primary index} defined over the primary key. Edge types do not require primary keys and there are no indexes on edges.  
It is also possible to declare secondary indexes on vertex attributes. There are no requirements on uniqueness or nullability on secondary index attributes.

Within a graph, to uniquely identify a vertex, we need to specify the tuple $\langle\mbox{\emph{type,primary-key}}\rangle$. Using the type, 
we can identify the relevant primary index and then retrieve the vertex by using the primary key in the index. Edges can't be identified 
directly except through the vertices to whom they are attached.  An edge is uniquely identified by the tuple 
$\langle\mbox{\emph{source-vertex,edge-type,destination-vertex}}\rangle$. This implies that given two vertexes, there can only be a single 
edge of a given type.

In terms of APIs we support the usual CRUD APIs on objects like vertices, edges, types and graphs.  We divide the APIs into two
classes: \emph{control plane} APIs which manipulate bulk objects like graphs and types and \emph{data plane} APIs which manipulate
fine grained objects like vertices and edges.  
In addition, we expose a set of transaction APIs: \textsc{CreateTransaction}, \textsc{CommitTransaction} and \textsc{AbortTransaction}.  
The \textsc{CreateTransaction} API creates a transaction object which can be used to group multiple data plane operations into a single 
atomic transaction. 
If a transaction is not specified for a data plane operation like \textsc{CreateVertex}, a transaction is implicitly created 
for that operation and committed at the end of the call. Unlike data plane operations, control plane operations cannot be grouped
under a transaction.  Each control plane operation executes under its own transaction. 

\subsection{Catalog}
A1 roots all data structure in the \emph{catalog}. It is a system data structure which returns handles to objects like tenants, graphs,
types, indexes, BTrees etc.  The catalog is fundamentally a key-value store where the key is the name of the object and the value is a
pointer to all the data needed to access the object.  For example in the case of a BTree, the catalog maps the name of the BTree to the
FaRM address of the root node of the BTree.  Once we have the root node of the BTree, we create an in-memory object called a BTree {\em
  proxy} which allows us to lookup/manipulate the BTree contents.  The catalog itself is stored in FaRM and hence materializing a proxy from
the BTree name can be an expensive operation ---it involves multiple remote reads to map the name to the root node and then potentially
reading the root node itself for any BTree metadata.  To reduce load on the catalog and as well as avoid remote reads in materializing
proxies, we cache proxies in memory once they are materialized.
Once cached, data plane operations like \textsc{CreateVertex} can use them without incurring the overhead of looking up the catalog separately. The cache has a fixed TTL to ensure that we don't use stale proxies.  When the TTL expires, the cache checks
if the underlying object has changed: if it has then we refresh the proxy, if it hasn't then we simple extend the TTL and continue to use the proxy.


\begin{figure}[htbp]
    \centering
    \includegraphics[width=0.8\columnwidth]{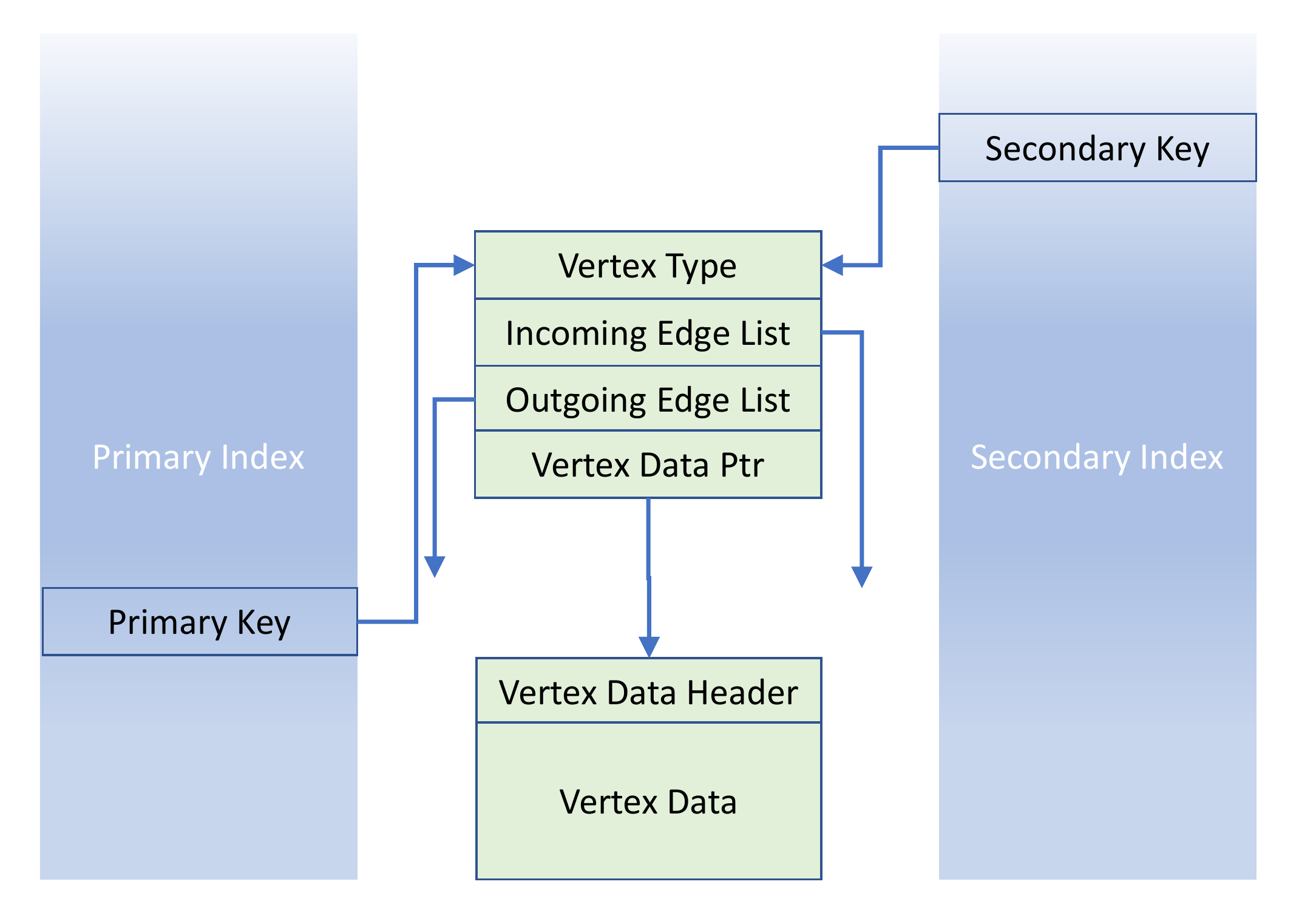}
    \caption{Vertex and primary index}
    \label{fig:index-and-vertex}
\end{figure}

\subsection{Vertices and Edges}

The storage format for vertices and edges are dictated by how they are accessed. For a vertex, we can look it up using either an index or
through edge traversal. Since traversal queries are much more frequent than index lookups, we optimize for that.  A vertex is
stored as two FaRM objects: a header object and a data object as shown in Figure \ref{fig:index-and-vertex}.  The vertex header contains the type of the vertex,
pointers to data structures that hold edges associated with the vertex, and a pointer to the data associated with the vertex.  
As the vertex is updated with  new edges or new  data, the header content changes, but the pointer to the header
itself remains unchanged.  We call this pointer the {\em vertex pointer}.  The vertex data is stored in a separate variable length object and serialized
in Bond binary format.  Since the data for a vertex is always schematized, the data representation is very compact and efficient to deserialize.  Since vertex data and
header are looked up together most of the time, we use locality to store both of them in the same region.

Looking up a vertex from its primary key is a multi-step process.  First, we look up the vertex pointer (address of the vertex header) from the index which is a BTree. We cache internal BTree nodes heavily\cite{farm-nsdi} and in most cases 
this lookup requires one RDMA read rather than $\mathcal{O}(\log(n))$.  Once the vertex pointer is found, we need two consecutive RDMA reads to read the header and then the actual data.  
If the vertex is being read during a traversal, then we can bypass the index lookup and we need only two consecutive reads.  

\begin{figure}[htbp]
    \centering
    \includegraphics[width=\columnwidth]{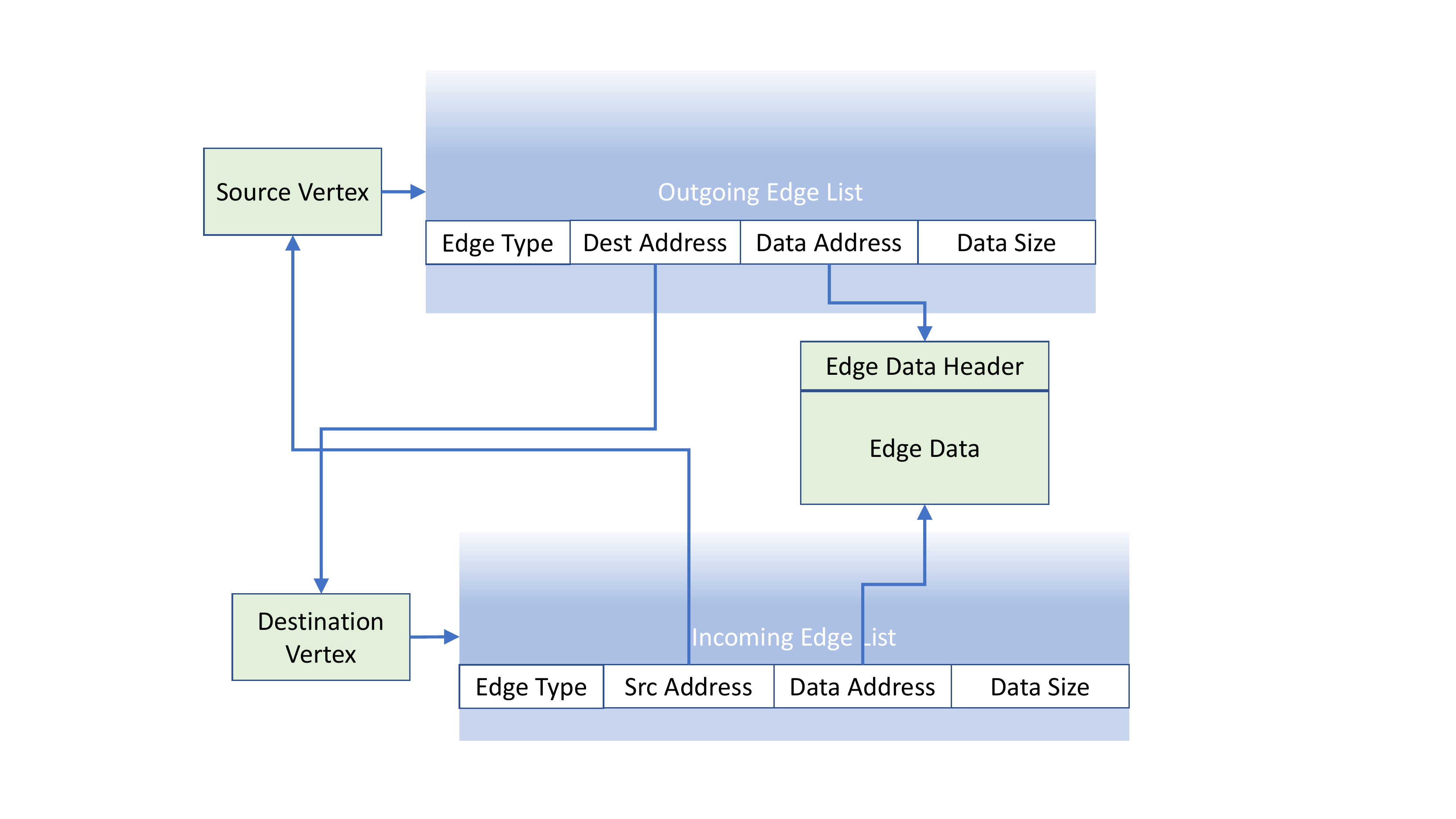}
    \caption{Vertex, edge lists and half-edge}
    \label{fig:edge}
\end{figure}

Unlike vertexes which can be uniquely identified by the vertex pointer, edges are not stored in a single unique FaRM object.  This is
again dictated by how edges are used inside A1 queries.  Given an edge $e$ from vertex $v_1$ to vertex $v_2$, if we delete $v_2$ we'd like
to ensure that the edge $e$ pointing from $v_1$ does not remain dangling.  To achieve this, we store the edge as a 3 part object as shown in Figure \ref{fig:edge}.
First we associate two {\em edge lists} with each vertex: an incoming edge list and an outgoing edge list.  The edge $e$ appears twice:
once as an entry in the outgoing edge list of $v_1$ and once in the incoming edge list for $v_2$.  We call the entry that appears on the
edge list a {\em half-edge}.  The outgoing half-edge for $v_1$ consists of the tuple $\langle$edge type, $v_2$ pointer, data
pointer$\rangle$ while the incoming half-edge for $v_2$ consists of the tuple $\langle$edge type, $v_1$ pointer, data pointer$\rangle$.
The data pointer field points to a FaRM object that holds the data associated with the edge.  In this example, if we delete $v_2$, then
by inspecting its incoming edge list, we know that there is an edge pointing to it from $v_1$ and we can go to $v_1$ and delete the entry 
in $v_1$ as well.  The edge list data structure needs to satisfy a few constraints ---first, since a single vertex can hold millions of edges
associated with it, the data structure needs to be scalable.  Next, given an edge characterized by the source vertex, destination vertex
and edge type, we should be able to lookup/insert/delete the edge quickly.

To satisfy these requirements, we actually use two different implementations of the edge list.  For small number of half-edges, all
half-edges are stored as an unordered list in a single FaRM object of variable length.  As the number of edges increase, we resize the 
FaRM object in a geometric progression until we reach around 1000 edges. For vertexes with more than 1000 edges, we store the edges in a global
BTree where the key is the tuple $\langle$src vertex pointer, edge type, dest vertex pointer$\rangle$ and the value is the edge data
pointer.  As long as the half-edges are stored in a single FaRM object, we use locality to locate that FaRM object with the associated
vertex.  Empirically we have found that for our current use cases, 99.9\% of the vertexes contain fewer than 1000 edges.  Given this edge
layout, once a vertex is read, enumerating its edges requires just one extra read as long as the number of edges is small.  Due to locality, this read is often simply a local memory access.

Although we take pains to allocate vertices and edge lists together using locality, we do not attempt to enforce locality between different vertices.  In case of immutable or slowly changing graphs, it is possible to run offline jobs to pre-partition a graph so that vertices connected together end up close to each other.  But we have avoided going down this route since it imposes considerable burden on our customers to do the offline graph partitioning. Also as updates happen, the original partitioning may no longer make sense.  Instead we believe it is the responsibility of the database to simplify the application developer's experience and provide acceptable performance. We currently place vertices randomly across the whole cluster and use locality to push query execution to where data resides.  Looking at sample query executions, we have found that this strategy can be highly effective (95\% local reads) and we will discuss more in section \ref{sec:performance}.

\subsection{Asynchronous Workflows}
Recall that APIs like \textsc{DeleteGraph} or \textsc{DeleteType} are asynchronous.  For example, calling \textsc{DeleteGraph} transitions the graph from 
state \textsc{Active} to \textsc{Deleting}, but the storage and resources associated with the graph is not freed 
synchronously.  Instead an asynchronous workflow is kicked off which deletes all the resources associated with the graph and finally frees
the graph itself. Before a graph can be deleted, all types associated with the graph is deleted.  For a type to be deleted, we delete all
the indices associated with the type: both primary and secondary. When the primary index is deleted, we delete the vertices at the same
time.

The asynchronous workflows run within the A1 process using what we call a \emph{Task} execution framework.  Tasks are units of work that can be scheduled to execute in future: tasks are enqueued on a global queue that is stored in FaRM.  We have a pool of worker threads on every backend machine that look for pending tasks and work on them. Since tasks are globally visible, any single task may be worked on any backend machine in the cluster.  The worker threads are stateless and they save their execution state in FaRM itself.  Once a task is scheduled, it is picked up by a worker thread.  If the thread can finish the task immediately, the task is completed and deleted.  Alternatively, if the task is bigger, the worker may reschedule the task to run in future or spawn more tasks to parallelize the execution.  This is the pattern we follow in the \textsc{DeleteGraph} workflow: the \textsc{DeleteGraph} API call simply creates a task.  When this task is executed by a worker, it spawns more tasks to delete all the types in the graph and waits for all those tasks to complete.  The \textsc{DeleteType} tasks in turn execute for a long time since each type needs to delete all the vertices, edges and indexes associated with the type.  Using this framework, we are able to harness the entire cluster's resources to execute long running workflows.  To ensure that the workflows do not interfere with real-time workload, the worker threads run at a low priority.


\subsection{Query Execution}
A1 workloads are dominated by large read-only queries which access thousands of vertices, and small updates that read and write a handful
of vertices. When a query/update operation arrives at the frontend, it is by default routed to a random backend machine in the cluster. There
are cases where more complex routing is required and we will discuss it later in the section.  When an update operation arrives at a
backend machine, that machine becomes the transaction coordinator for the operation. All read and writes are executed on that machine using
RDMA to access remote data. Writes are made durable during transaction commit using one-sided RDMA  writes.   Queries are executed a little
differently: the backend machine where the query arrived first, is designated as the \textit{coordinator} for that query, which drives the
execution of the query, but the bulk of the query execution work is distributed across the cluster.  We designate the machines where
query execution happens at the instigation of the coordinator as \textit{workers}.

To understand the A1 query language and its execution, let's take as an example, a knowledge graph of films, actors and directors.  If an actor appears in a film, then the film is connected to the actor with an outgoing edge of type \textit{film.actor}.  Similarly the director and the film is connected with an edge of type \textit{director.film}. The A1
query language known as A1QL is similar to MQL: the Metaweb Query Language \cite{mql}.  Let's consider the two-hop query that asks for all actors that
worked with Steven Spielberg. in A1QL, the query is written as shown in Fig.\ref{fig:2-hop-query}.

\begin{figure}[htbp]
\begin{verbatim}
{ "id" : "steven.spielberg",
  "_out_edge" : { "_type" : "film.director",
      "_vertex" : {
          "_out_edge" : { "_type" : "film.actor",
              "_vertex" : {
                  "_select" : ["*"] 
              }}}}}
\end{verbatim}
\caption{A1 query to retrieve all actors that have worked with Steven Spielberg}
\label{fig:2-hop-query}
\end{figure}

Every A1 query is a JSON document with each level of nested JSON struct describing a step in the traversal with the starting point at
the top level document.  In this query, the top level struct specifies the starting vertex as the vertex with primary key
\textit{steven.spielberg} and we use the \textit{id} field to look up the director from the primary index.  The next level specifies that we
should traverse an outgoing edge (\textit{\_out\_edge}) of type \textit{film.director} to a film. The next level describes that we
should traverse out on an edge of type \textit{film.actor} to from the film to arrive at the actor vertex.  At the last level, the \texttt{select(*)}
clause indicates that we should return all values.

\begin{figure}[htbp]
  \centering
    \includegraphics[width=0.8\columnwidth]{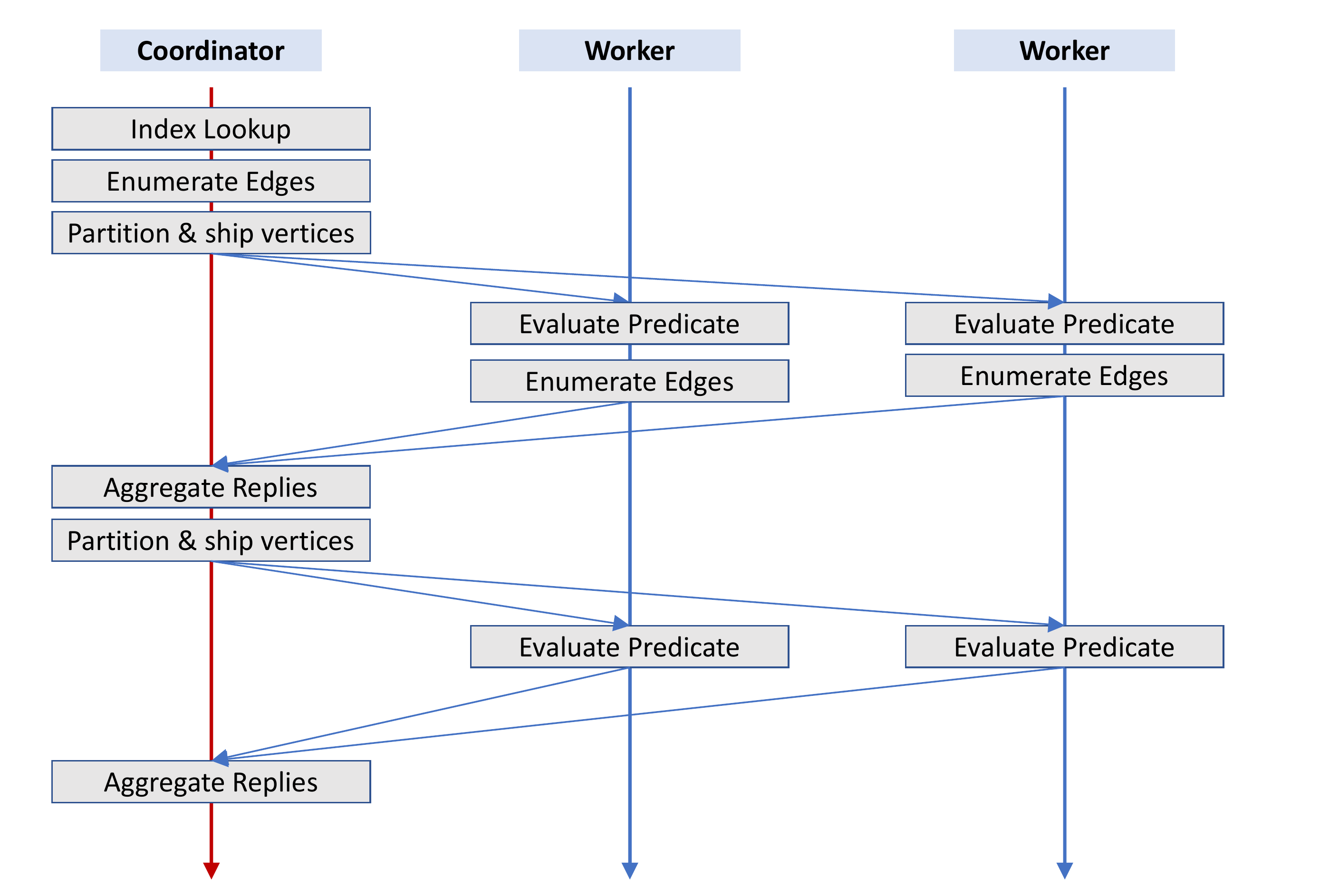}  
\caption{Physical A1 query execution.  The query is to retrieve all
  actors that have worked with Steven Spielberg (Fig.\ref{fig:2-hop-query}.}
\label{fig:query-exec}
\end{figure}

Now let us see how the example query from Figure  \ref{fig:2-hop-query} is executed.  The query coordinator parses the query to derive a
logical plan and then generates a physical plan.
A1 doesn't have a true query optimizer: most of the queries submitted to A1 are straightforward and executed without any optimization. In A1QL the user can supply some optional optimization hints.  If they are supplied, then they are used in creating the physical  execution plan from the logical plan. Building a true optimizer is currently work in progress.  Queries are built on top of a few basic operators like
index scan, predicate evaluation against a vertex/edge data and edge enumeration for a given vertex. The step by step query execution is
shown in Figure \ref{fig:query-exec}. In this execution, the coordinator starts by instantiating a transaction and choosing the transaction
timestamp as the version which will be used for all snapshot reads. Next it does an index lookup to locate the \textit{Steven Spielberg}
vertex and then from the vertex, enumerates  all neighboring half-edges of type \textit{film.director}.

The edge enumeration gives the coordinator a list of vertex pointers for all the films (see Figure \ref{fig:edge}). In the next step, we need to
look up all the actor edges from those film vertexes. Since the edge list is co-located with the vertex, it is more efficient to execute the
task of edge enumeration at the actual location of the vertexes. Therefore the coordinator maps the vertex pointers to the
physical hosts which are the primary storage hosts for the corresponding vertex.  Mapping pointers to physical hosts is a local
metadata operation with no remote accesses.   Operators like predicate evaluation and edge enumeration are shipped to the
machine hosting the vertex via RPC so that it can be evaluated without invoking a remote read. When we have multiple vertex operators to be
processed at the same machine, we batch the operators together per machine to reduce the number of RPCs.  Although query shipping is the
norm, if the number of vertexes operators to be shipped are too small we avoid the RPC overhead by evaluating the operators locally using RDMA reads.

Each worker receives RPCs from the coordinator and instantiates a new read-only transaction at the timestamp chosen by the query
coordinator.  This ensures that all the query reads form a consistent global snapshot across the entire distributed graph. The typical
operator that executes in a worker are predicate evaluation which applies predicates against vertex data and edge enumeration for the
vertex.  Note that both of these operations do not require any remote reads assuming locality applies.  During edge enumeration, any edge
predicate is also applied.  Once edge enumeration is finished, the results are a set of vertex pointers for the next hop of the
traversal, i.e. a set of actor vertices.  These vertices are shipped back to the coordinator where they are aggregated, duplicates removed
and repartitioned by pointer address to run the next phase of the traversal.  Once the whole query completes, the results are aggregated
by the coordinator and returned to the client.  Since we keep the entire state of the query in the memory of the coordinator, we are 
vulnerable to queries which require a working set bigger than the coordinator's available memory.  Implementing disk spill for such a 
case is infeasible since our goal is to be a low-latency system.  Currently we simply fast-fail queries whose working set grows too 
large ---in future we plan to dedicate regions in the cluster for spilling intermediate query results.  Fast-fail is  
an acceptable option since very large queries typically will not be finished within its time budget anyway.  

If the final result set is too large to return in a single RPC, the coordinator does not return the full result set and instead return partial results and a continuation
token.  Rest of the results are cached in the coordinator and can be retrieved by the client by supplying the continuation token in the next request.  The
continuation token encodes the coordinator host's identity in it. When a request for result retrieval using a continuation token is received
by a frontend, the frontend decodes the coordinator's identity and forwards it to the correct machine so that rest of the results can be
returned. The coordinator caches the results only for a limited time (typically 60 seconds) to conserve resources ---the client is expected
to retrieve all the results in that time.  If the cache times out or the coordinator crashes before all the results are returned, the
client is expected to restart the query.  Since our typical query execution lifetime is measured in less than a second, this is not a
big concern.

\section{Disaster Recovery}\label{sec:replication}
Although FaRM replicates data in memory 3-ways, there are situations where data can be lost such as power loss to an entire datacenter or coordinated failure of 3 replicas.   Therefore any system built on FaRM needs to have a {\em disaster recovery} plan.   A1 implements disaster recovery by replicating all data asynchronously to a durable key-value store known as ObjectStore which is used by Bing.
 We will not go into details of ObjectStore except to say that it supports the abstraction of tables with each table containing a large number of key-value pairs.  Both keys and values are schematized using Bond.  Writes to ObjectStore made durable by replicating every write 3-ways into durable store.

Since the replication from A1 to ObjectStore is asynchronous, in the event of a disaster, ObjectStore may not contain all the writes committed to A1.  To deal with this data loss, our recovery scheme supports two types of recovery: {\em consistent recovery} and {\em best-effort recovery}.  In consistent recovery we recover the database to the most up to date transactionally consistent snapshot that exists in ObjectStore.  With best-effort recovery, we do not guarantee that the recovered state of A1 will be transactionally consistent, but the database itself will be internally consistent.  Let's take an example to illustrate this.  Suppose we have a single transaction in A1 that adds two vertexes, A and B and an edge from A to B.  We take a few example scenarios:
\begin{itemize}
\item We succeed in replicating A and B, but the edge is is not replicated.  In that case after consistent recovery, A1 will not contain any of A or B or the edge.   On the other hand, best effort recovery will recover both A and B, but there will be no edge between them.
\item We succeed in replicating A and the edge, but not B.  Again, consistent recovery will treat this as a partial transaction and will ignore the edge and A.  Best effort recovery will recover A and notice that the other end of the edge B is missing and will not recover the edge.  Therefore the database will be internally consistent ---no dangling edges, but not transactionally consistent.
\end{itemize}
Best-effort recovery therefore always recovers the database to a state which is at least as up to date as consistent recovery and in almost all practical cases, to a more up to date state.

For every graph, we create two tables in ObjectStore to durably store the data: the vertex table stores all the vertexes regardless of the vertex type, while the edge table stores all the edges.  When an update request arrives at A1, we apply the update to A1 and also insert a log entry for the update to a {\em replication log} transactionally.  The replication log is itself stored in FaRM with the usual 3 copy in-memory replication guarantee.  As soon as the update transaction commits, we attempt to replicate the update in replication log to ObjectStore synchronously with the customer request.  If the replication effort succeeds, then we delete the log entry and acknowledge success to the client.  If the replication effort fails, we have an asynchronous replication sweeper process that scans the replication log in FIFO order and flushes the unreplicated entries to ObjectStore and if successful, delete the entry.  We closely monitor the age of entries in the replication log to make sure we do not have too many entries in there ---in ideal case, the replication log should be empty except for ongoing update transaction entries.   In case of a disaster, the entries in the replication log which were not replicated to ObjectStore synchronously are the ones which will be permanently lost.



When we replicate entries from replication log to ObjectStore, we need to make sure that entries are applied in the same order as the transaction order in A1, i.e. if we stored value $v_1$ in vertex $V$ and then store value $v_2$ in V, then regardless of delays or failures in the replication pipeline, eventually when all updates are flushed from the replication log, ObjectStore must reflect value $v_2$ as the final value.  This is achieved differently for consistent recovery and best-effort recovery.   Recall that in FaRM, every write transaction is assigned a global commit timestamp which imposes a global order among all transactions that occurred in the system. In best effort recovery, every row in the vertex or edge table has a timestamp field which corresponds to the timestamp of the FaRM transaction responsible for that update.  When a new update comes in, we compare the timestamp of the existing row in ObjectStore with the update's timestamp.  If the update's timestamp is newer, then the update is a later transaction and we store the update into the row.   On the other hand, if the update is older than the existing content of the ObjectStore table row, then this update is a stale update and we can discard it.  For create operations, we unconditionally create the new row, while for delete operations we create a tombstone row with the delete timestamp.   The tombstone entry is removed either when the row is recreated with a newer timestamp or by an offline garbage collection process which removes all tombstones older than a week.  For the sake of efficiency, we do not explicitly do a read-modify-write to implement this protocol: ObjectStore exposes a native API that accepts a timestamp version and achieves this is a single roundtrip.  Note that this update process is idempotent: if a replication log entry is flushed multiple times, the outcome is not changed.

Consistent recovery works a little differently: in this case we treat ObjectStore as a versioned datastore.  Instead of just storing $\langle$key$\rightarrow$value$\rangle$ rows in the ObjectStore table, we augment the key by the transaction timestamp version to get the row $\langle$(key,timestamp)$\rightarrow$value$\rangle$.   Since ObjectStore supports iterating over keys in sorted order, given a key, it is easy to find all versions of that key or the latest version of the key.  When an update comes in with a given timestamp, we always insert it into ObjectStore.  For deletes, a tombstone entry is inserted.  Again, this protocol is idempotent.  To recover to a consistent snapshot from this durable versioned store, we need to find a timestamp value below which all updates in A1 are also reflected in ObjectStore.  To do this, A1 continually monitors the timestamp of the oldest unreplicated entry in the replication log ($t_R$) and stores this value to ObjectStore durably.  Clearly when $t_R$ is made durable in ObjectStore,  all writes that have timestamp smaller than $t_R$ are also durable in ObjectStore.  On recovery, we read the value of $t_R$ and recover using the snapshot corresponding to this timestamp.


\section{A1 in Bing}\label{sec:applications}
In this section we will first look at Bing's use of A1 and then focus on our experience bringing A1 into production use.

A1 is designed to be a general purpose graph database and there are
multiple applications in Bing that runs on top of it. In this paper, we focus on a single use case: knowledge graph.  We have already
encountered a few knowledge graph serving example scenarios. The
knowledge graph is generated once a day by a large scale map-reduce
job.  There are real-time updates to the knowledge graph as well. The original Bing knowledge graph stack was a custom-built system with
immutable storage and regular key-value store. This prohibited real-time updates and could not handle more complex queries within the 
latency constraint. A1 addresses both of these shortcomings and increases the overall flexibility of the system.

In A1, the knowledge graph is designed with a semi-structured data model.  All entities whether they are
films, actors, books or cities are modeled as a single type of vertex named \textit{entity} with all attributes stored as a key-value map.  This is a choice necessitated
by the fact that the number of different types of entities in a 
knowledge graph is vast (tens of thousands) and their attributes are constantly
changing because we add more and more information to entities.  On the
other hand, we strongly type the edges since the edge types are
typically fixed and there is little data associated with edges.
In practice, we have found that weak typing of vertices do not lead to
significant query slowdowns while enabling more flexible data modeling.
Since A1 storage is expensive, only queryable attributes of an entity
are stored in A1 while non-queryable attributes like image data are
stored elsewhere.

Bing receives human generated queries like ``Tom Hanks and Meg Ryan
movies'' which are translated to A1QL queries.  The translation step
is non-trivial since it requires us to map strings like ``Tom Hanks'',
``Thomas Hanks'' or even just ``Hanks'' to the unique actor entity
\textit{Tom Hanks} that we all know and love.  We will not go into the
complexities of query cleaning and query generation here in this paper.
The results of the queries are joined with data from other sources to
render the final page view.  For example in this query if we return
the vertex corresponding to the film \textit{You've Got Mail}, the
rendering pipeline pulls together image data for the movie (e.g. movie
poster) and generates the final page.  Overall, A1 improves the
average latency of the knowledge serving system by 3.6X and enables
significantly more complex queries.

\subsection{RDMA in the Data Center}
RDMA originated in rack scale systems and is a difficult protocol to work with in the large data center networks. 
Since RoCEv2 doesn't come with its own congestion control, we use DCQCN \cite{dcqcn} to enforce our own congestion 
control and fairness. We handle a lot of the protocol level instability by defensive programming around communication problems.  FaRM is able to recover very quickly from any host/network level failures which ensure that users do not notice networking hiccups.  In addition to RDMA \textsc{Read} and \textsc{Write}, we also make heavy use of RDMA unreliable datagrams (UD) for clock synchronization and leases. In general we 
have been able to achieve latencies less than 10 microseconds within a single rack and less that 20 microseconds across racks with oversubscribed network links.  

\subsection{Opacity and Multiversioning}
In building A1, we enhanced FARM from the version described in \cite{farm-nsdi, farm-sosp} (denoted FARMv1) with several features into what we will call FARMv2. 


The isolation guarantee provided by FaRMv1 transactions is
serializability, but combining serializability with optimistic
concurrency control can lead to certain well known problems. For
example, consider two transactions: $T_1$ reading a linked list
consisting of two items $A\rightarrow B$, and $T_2$ deleting $B$ from
the list concurrently.  Suppose the execution interleaving of the transactions is
the following:
\begin{enumerate}
\item $T_1$ reads $A$ and gets the pointer to $B$.
\item $T_2$ deletes $B$ and commits.
\item $T_1$ dereferences the pointer to $B$ which is now pointing to
  invalid memory. The application reads the invalid content of $B$ and
  panics.
\end{enumerate}
Since executions of $T_1$ and $T_2$ are serializable, $T_1$ will abort
once it attempts to commit, but even before that, the application will
conclude erroneously at the last step above that the data it has read
is corrupt.  The solution to this problem is known as the
\textit{opacity}\cite{opacity} property which guarantees that even transactions that
will eventually abort (e.g. $T_1$) are serializable with respect to
committed transactions (e.g. $T_2$) and hence will no longer cause
application inconsistencies at runtime.

Optimistic concurrency control can often lead to high abort rates for large transactions.  A1 is an OLTP system
which combines small update transactions (touching a handful
vertices/edges at most) with much larger read-only queries which can
read many thousands of vertexes in a single query. Since optimistic
concurrency control does not acquire read locks, the large queries are
susceptible to conflict with updates and hence abort frequently.

FaRMv2 solves both of these problems by introducing a global clock
which provides read and write timestamps for all transactions.  These
timestamps provide a global serialization order for all
transactions. In addition FaRMv2 implements MVCC (multi-version
concurrency control) which ensures that read-only transactions can run
conflict-free with update transactions.  For details on the
implementation, we refer the reader to FaRMv2 paper\cite{farm-sosp}.  The fact that
all transactions can be ordered globally using their write timestamp
is also used in our disaster recovery solution which we
discussed in section \ref{sec:replication}.

\subsection{Fast Restart}
Data stored in FaRM can be made durable\cite{farm-sosp} using SSD for storage and using non-volatile RAM (NVRAM) for transaction log durability.  But since A1 runs on commodity machines with no NVRAM, the
durability problem needs to be solved differently.  There are two
different durability problems that we address.  First, if we lose
power to the entire data center, clearly all data in memory in A1 will
be lost.  We consider this as a disaster scenario and implement
disaster recovery, which was described in section \ref{sec:replication}.

A software outage in 3 machines across 3 failure domains can occur during
deployment or due to a bug. In the case that these 3 machines hosts the
3 replicas of a single region, a total loss of that region will occur. This 
implies losing parts of the graph or index, and should be considered catastrophic.

We protect against this by implementing a feature known as \textit{fast
  restart}. In FaRM, the memory where the regions are allocated do
not belong to the FaRM process itself: instead we use a kernel driver
known as PyCo, which grabs large contiguous physical memory segments
at boot time. When the FaRM process starts, it maps the memory
segments from the driver to its own address space and allocates
regions there.  Therefore, if the FaRM process crashes unexpectedly,
or restarts, the region data is still available in the driver's
address space and the restarted process can grab them again. Note that
fast restart doesn't protect against the machine crash or power cycle
because in that case the machine will reboot and the state held in the
driver's memory will be lost. In FaRMv1, only the data regions were
stored in PyCo memory. As part of fast restart, we moved all data
needed to correctly recover after a process crash to PyCo memory
---this includes region allocation metadata and transaction
logs. Recall that the configuration manager (CM) is responsible for
determining which regions are hosted in which machines.  In case of
any machine failure, if the CM detects that all replicas for any
region has been lost, it pauses the whole system and all transactions
are halted. In case of accidental A1/FaRM process crash, our
deployment infrastructure automatically restarts the process.
Therefore the CM waits to see if the failed process or processes will
come back to life and if they come back it initiates recovery of that
region's data including all blocked transactions.  Overall, fast restart has cut down the downtime for A1
cluster by an order of magnitude.

\section{Performance Evaluation}\label{sec:performance}
To evaluate the performance of A1 experimentally, we use a graph
consisting of 3.7 billion vertices and 6.2 billion edges, which is
generated from the film and entertainment knowledge base containing
22.9 billion RDF triples with 3.7 billion entities. Graph vertices
represent entities and have several attributes, and edges do not have
data attributes. On average every vertex had a payload of 220 bytes. Although the average vertex degree is small, the skew
in vertex degree distribution is very large and some vertices have
degrees larger than ten million.

We use a cluster of 245 machines and
measure end-to-end response time from a client in the same datacenter
as the A1 cluster. Every machine has two Intel E5-2673 v3 2.4 GHz
processors, 128GB RAM and Mellanox Connect-X Pro NIC with 40Gbps
bandwidth. A1 uses 80GB of the available RAM for storage. The total storage space available in the machines is 245*80GB/3 = 6.5TB --the factor of 3 is for 3x replication.  Our data occupies 3.2TB of the total available space. The machines are
distributed across 15 racks and four T1 switches connect the
racks. ToR (Top of the Rack) switches provides full bisection
bandwidth between machines in a single rack, while T1 switches use oversubscribed links between racks. Therefore, most of the
cross-machine traffic uses oversubscribed links. Vertices are
distributed at random across the machines, and therefore 99.6\%
(=244/245) of a vertex neighbors are on a remote machine.  We report
the average and P99 (the 99th percentile) latency for a few multi-hop
queries which represent various types of graph queries.


\begin{table}[htbp]
  
 \begin{tabular}{|l|l|}
   \hline
   Id & A1QL \\
   \hline
   \hline
   Q1
   &
   \begin{minipage}{3in}
   {\small
\begin{verbatim}
{ "id" : "steven.spielberg",
  "_out_edge" : { "_type" : "director.film",
      "_vertex" : {
          "_out_edge" : { "_type" : "film.actor",
              "_vertex" : {
                  "_select" : ["_count(*)"] }}}}}
 \end{verbatim}
 }
\end{minipage}
   \\
   \hline

   Q2
   &
   \begin{minipage}{3in}
{\small
\begin{verbatim}
{ "id" : "character.batman",
      "_out_edge" : { "_type" : "character.film",
          "_vertex" : {
              "_out_edge" : { "_type" : "film.performance",
      "_vertex" : {
          "str_str_map[character]" : "Batman",
          "_out_edge" : { "_type" : "performance.actor",
               "_vertex" : {
                   "_select" : ["_count(*)"] }}}}}}}
\end{verbatim}}
\end{minipage}
   \\
   \hline
   Q3
   &
   \begin{minipage}{3in}
{\small
\begin{verbatim}
{ "id" : "steven.spielberg",
  "_out_edge" : { "_type" : "director.film",
      "_vertex" : {  "_type" : "entity",
          "_select" : ["name[0]"],
          "_match" : [{
              "_out_edge" : { "_type" : "film.actor",
                  "_vertex" : { 
                      "id" : "tom.hanks"
                  }}},
              { "_out_edge" : { "_type" : "film.genre",
                  "_vertex" : {
                      "id" : "action"
                  }}}] }}}}
\end{verbatim}}
\end{minipage}
   \\
   \hline
   Q4
   &
   \begin{minipage}{3in}
{\small
\begin{verbatim}
{ "id" : "tom.hanks",
  "_out_edge" : { "_type" : "actor.film",
      "_vertex" : {
          "_out_edge" : { "_type" : "film.actor",
              "_vertex" : {
                  "_out_edge" : { "_type" : "actor.film",
                      "_vertex" : {
                          "_select" : ["_count(*)"] }}}}}}}
\end{verbatim}}
\end{minipage}
   \\
   \hline
 \end{tabular}
  \caption{Queries used to evaluate A1 performance}
  \label{tab:perf-queries}
\end{table}

We focus on  the following set of specific queries and see how the system performs.  The actual 
representation of the queries in A1QL is in Table\ref{tab:perf-queries}.
\begin{itemize}
\item   Q1: Count actors who have worked with Steven Spielberg.
\item   Q2: Count actors who have played Batman.
\item   Q3: Action movies with Steven Spielberg and Tom Hanks.
\item   Q4: Count number of films by actors who have worked with Tom Hanks.
\end{itemize}

\begin{figure}[htbp]
    \centering
    \includegraphics[width=0.8\columnwidth]{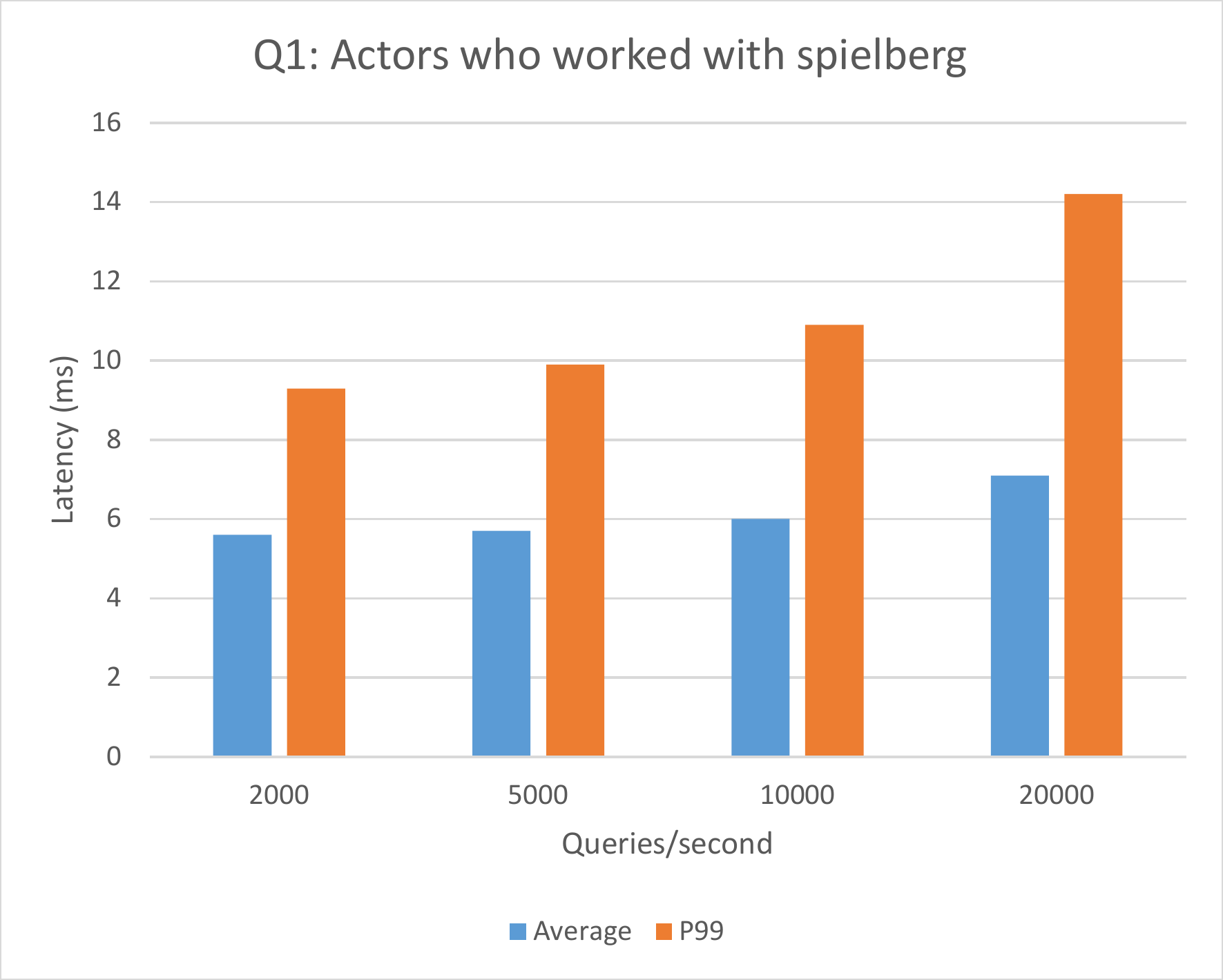}
    \caption{Average and P99 latency for Q1.}
    \label{fig:q1}
\end{figure}

The first query, Q1, asks for all actors that have worked with the
director Steven Spielberg. This translates to a simple 2-hop query
where we look up all films for Spielberg and then for those films find
all actors that have acted in them. Figure \ref{fig:q1} depicts the
average and P99 latency in for Q1, which reads a total of 49 vertices
in the first hop (films by Spielberg) and 1639 vertices in the second
hop (actors in those films). The total number of edges visited were
1785: this number is larger than the number of vertices since multiple
edges could point to the same end vertex. By parallelizing all these
reads across the cluster, we were able to complete this query in less
than 8ms on average and 14ms at p99 at 20000 queries/second.  Note the tight spread between the average and P99 latencies which is a consequence of the focus on latency for A1.

The total number of raw FARM objects read during the
query is 3443 out of which only 163 are remote. In other words, we
achieve more than 95\% local reads through query shipping to
workers. 
\begin{figure}[htbp]
    \centering
    \includegraphics[width=0.8\columnwidth]{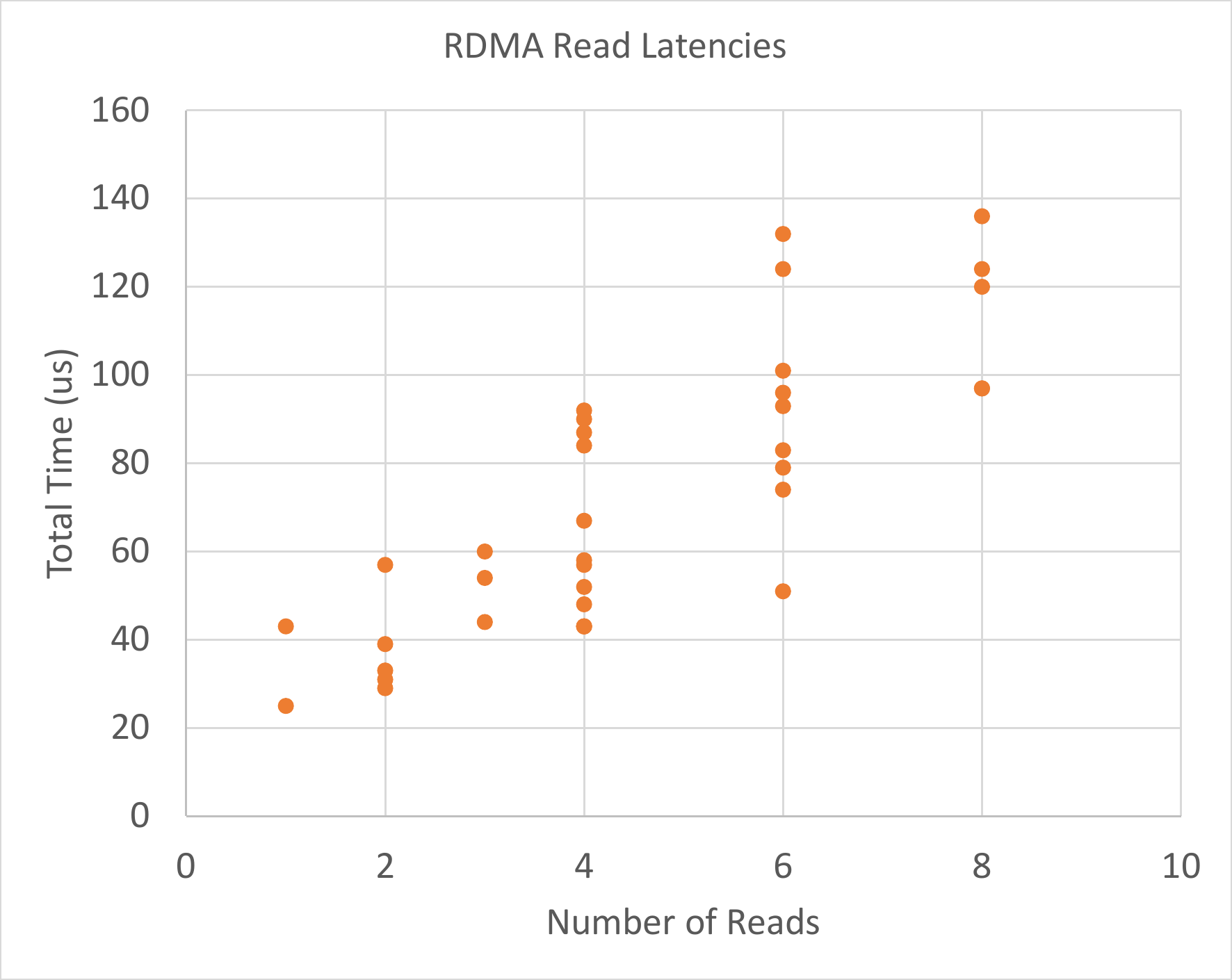}
    \caption{Total RDMA read latency (microseconds) for different number of read operations.}
    \label{fig:rdma-latency}
\end{figure}
Figure \ref{fig:rdma-latency} shows the distribution of RDMA latencies
as a function of number of reads done. Recall that we ship the vertex
predicate evaluation and edge enumeration to workers. So if a worker
lands with a bunch of vertices which are remote than it has to do one
or more RDMA reads to get all the data. Figure \ref{fig:rdma-latency}
shows the total time in microseconds doing RDMA reads vs the total
number of reads done and the trend is roughly linear. Average read
times for RDMA was 17us.  

\begin{figure}[htbp]
    \centering
    \includegraphics[width=0.8\columnwidth]{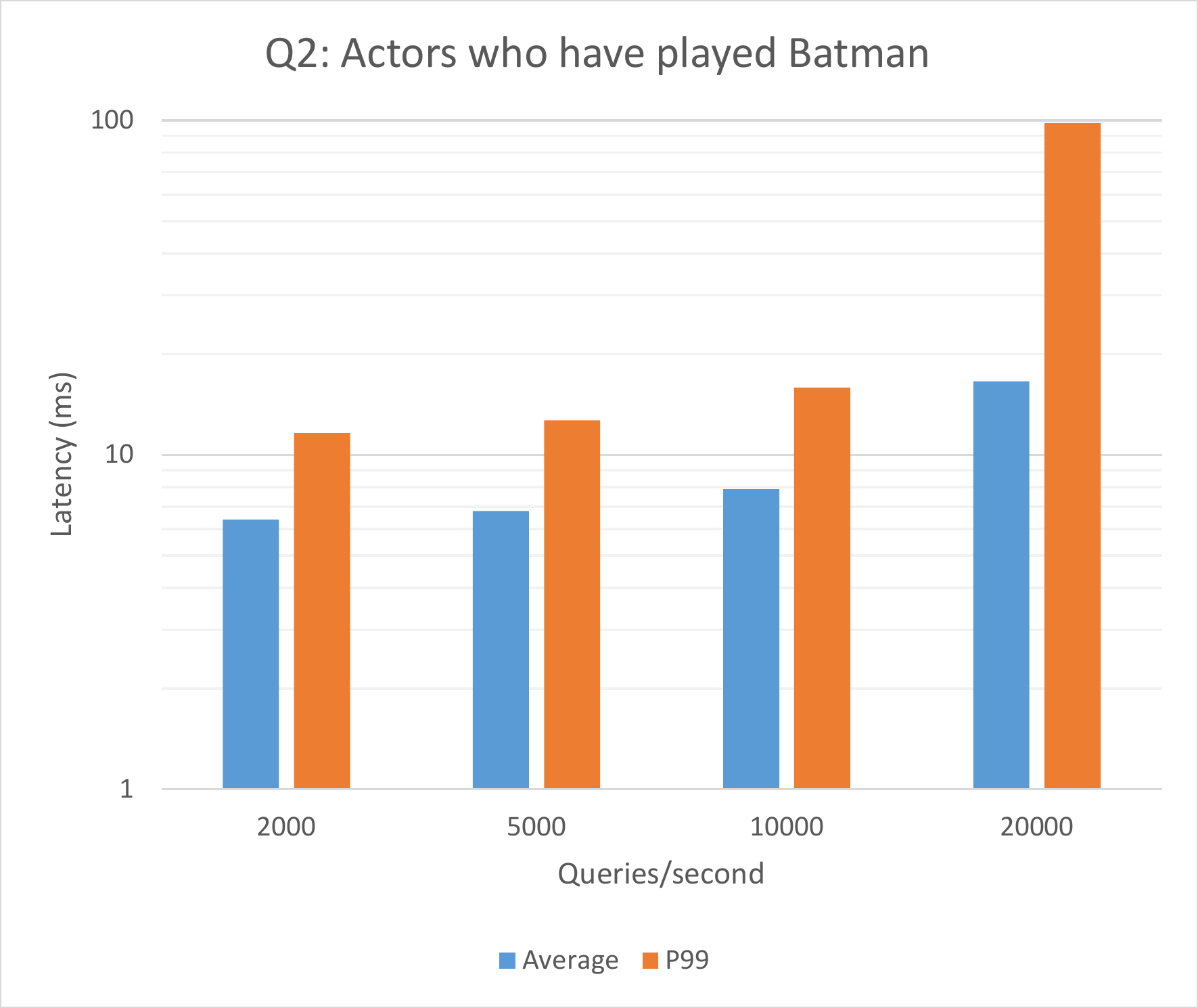}
    \caption{Average and P99 latency for Q2.}
    \label{fig:q2}
\end{figure}

Q2, is deceptively simple, but more complex
in its implementation: find all actors who have played Batman. This
maps to the following traversal where we first look up the entity
Batman and then all movies in which this
entity appears. For each of the movies, we look up the performances
of all actors and then filter those performances by name of the
character (Batman) and then the actor for that performance. This
translates to a three-hop query from character to film to performance
to actor (Figure \ref{fig:q2}.

\begin{figure}[htbp]
    \centering
    \includegraphics[width=0.8\columnwidth]{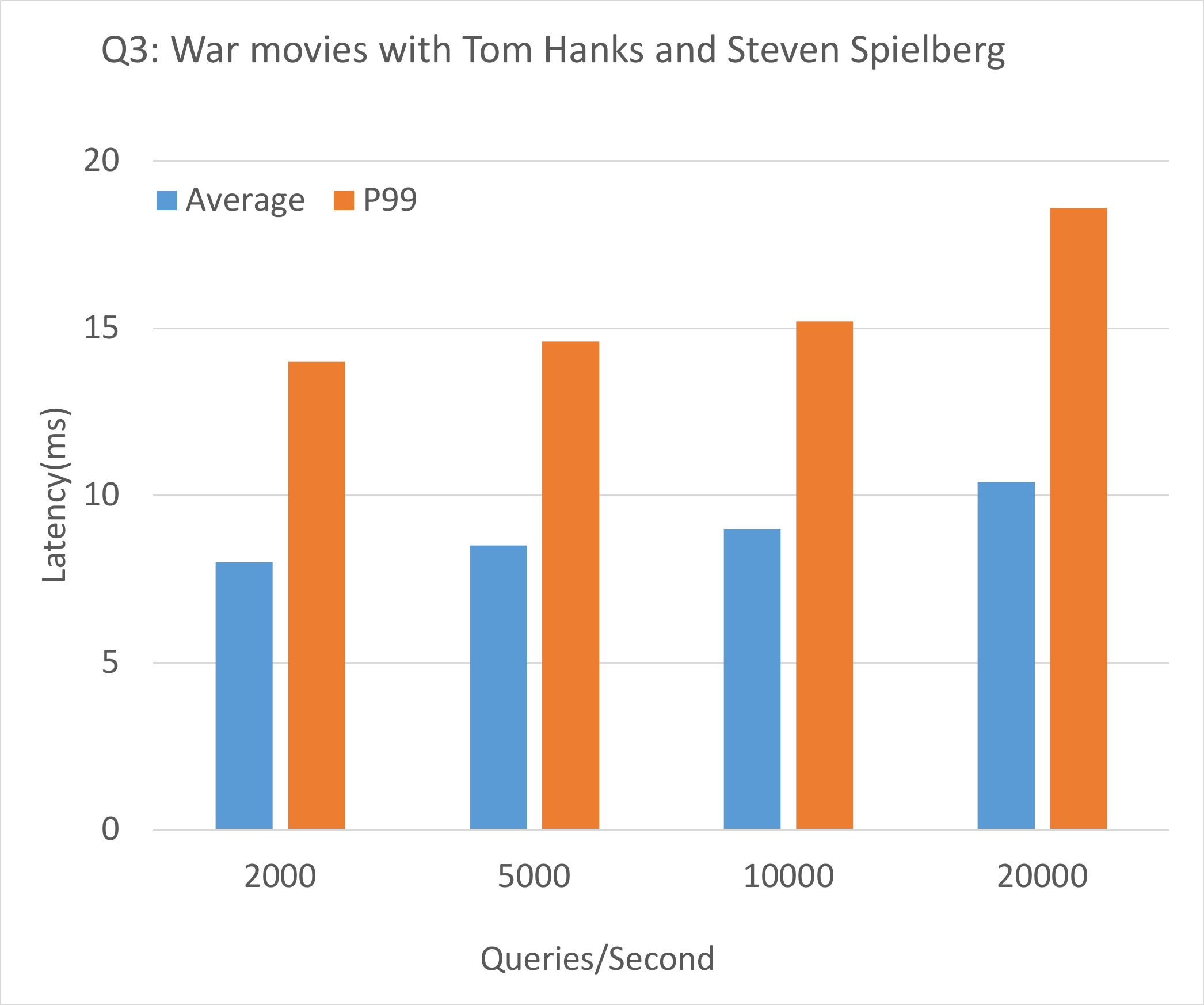}
    \caption{Average and P99 latency for Q3.}
    \label{fig:q3}
\end{figure}

Query Q3 (Figure \ref{fig:q3}) represents a more complex pattern of graph exploration. The query is to find all Spielberg
movies which belong to the War movie genre and stars Tom Hanks. Here
the graph pattern we are interested in is a star pattern where the
center is the movie and the movie is connected to three entities:
Spielberg as director, War as genre and Tom Hanks as an actor. A
similar query is to find all comedies starring both Ben Stiller and
Owen Wilson.

To  evaluate maximum throughput of the system, we carried out a test using Q4. 
For a given actor, Q4 finds all actors he/she has
worked with and finds films starring them. This maps to a three-hop
traversal query from actor to films to actors (co-stars) to their
films. The goal of Q4 was to stress the system by exploring a large number of vertexes rather than being a realistic user query. On average, Q4 accesses 24,312 vertices with 33ms latency for
throughput (1000 queries/second).  We pushed the cluster to 15,000
queries/second and at this throughput this query executes 365MM vertex
reads/second across the cluster, i.e. 1.49MM vertex reads per second
for every machine in the cluster.


\begin{figure}[htbp]
    \centering
    \includegraphics[width=0.8\columnwidth]{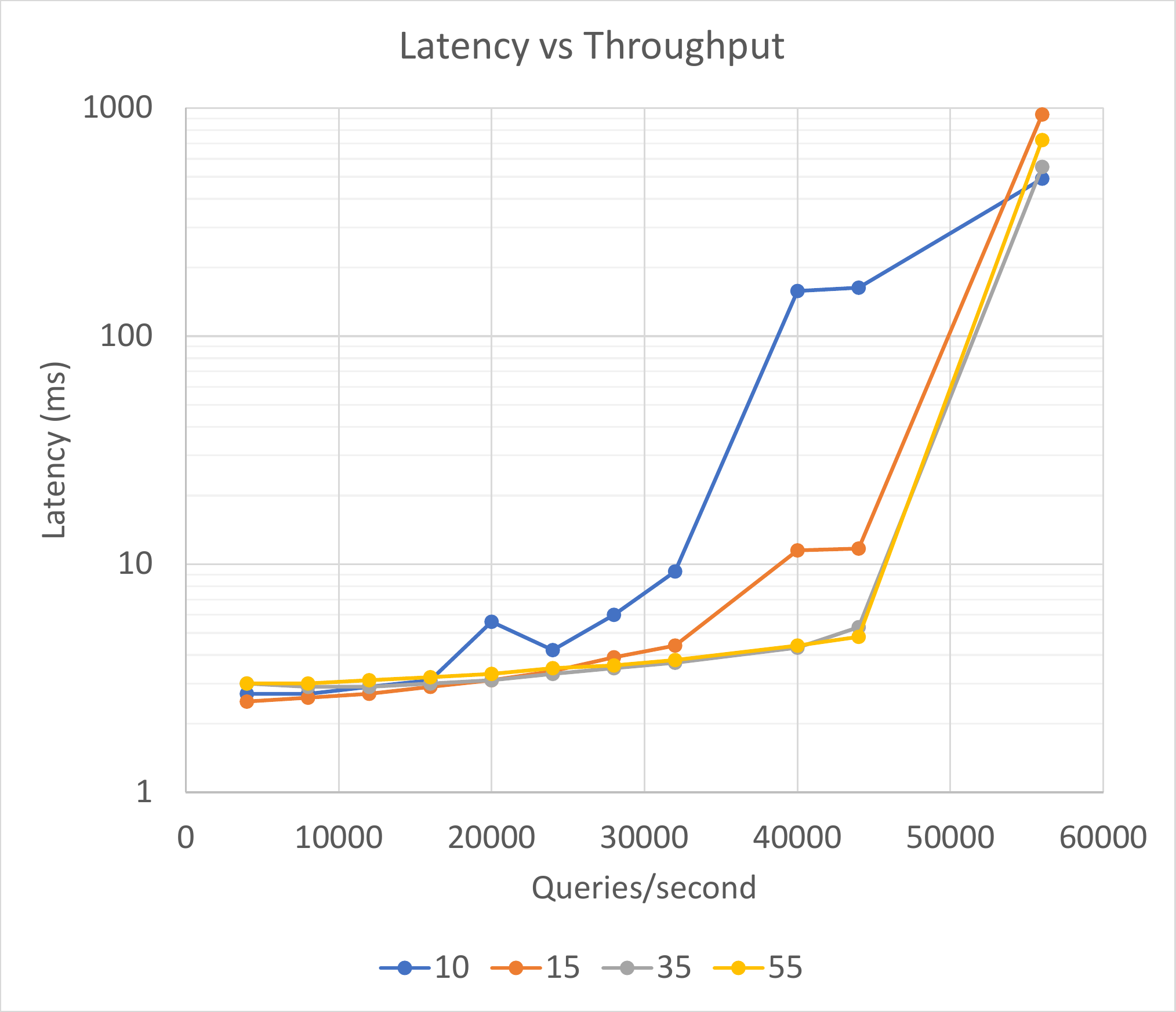}
    \caption{Latency vs throughput for different cluster sizes (10, 15, 35 and 55).}
    \label{fig:q-scale}
\end{figure}

Finally, to understand the scalability characteristics of A1, we created clusters of 10, 
15, 35, and 55 machines in the same network configuration as used for the larger cluster. 
We used a smaller dataset of 23 million vertices and 63 million edges. This dataset was distributed uniformly across the machines, and then 
ran a set of 2-hop queries. For each cluster, we measured the latency at different 
query loads, as shown in Figure \ref{fig:q-scale}.  As expected, the usable throughput (below a given latency) correlates to the
cluster size;  Latency of queries below the capacity threshold 
is mostly flat as the cluster grows. The clusters have the same network topology, so this is as expected. For larger clusters, the
expected benefit is not just scalability of throughput, but also capacity for bigger datasets.

\section{Related Work}
Graph databases are not new in the database world and in recent years,
they have been experiencing great interest in industry: some
notable efforts in the open source space include Neo4J\cite{neo4j},
Apache Tinkerpop\cite{tinkerpop} and DataStax Enterprise
Graph\cite{dse}, while AWS Neptune\cite{neptune} and
CosmosDb\cite{cosmosdb} are prominent cloud based offerings.  All of
these systems are disk based and apart from DataStax Enterprise Graph
and CosmosDb, none of them are distributed.  Traditional commercial
databases like Oracle and SQLServer also now support the graph data
model and associated query capabilities.

Graph data has been represented in various ways using RDF triples as well
as property graph like model.  RDF triples have been stored directly
in relational stores \cite{chong05} or stored in more efficient
columnar formats \cite{abadi07}.  Storing RDF data in relational
stores allows one to take advantage of the existing depth of the
relational technology.  Since A1 was built ground up as a new system
and the FaRM data model was highly conducive to building linked data
structure, we opted to go with a property graph model rather than RDF or relational.  Moreover, we
have found that most of our customers prefer the property graph model
in modeling their data.

Trinity\cite{trinity, trinity2} from Microsoft Research is the system
closest to A1 in terms of its use of in-memory storage and horizontal
scalability.  But compared to A1, Trinity lacks transactions and not
comparable in terms of performance.  Facebook's TAO\cite{fb-tao} and
Unicorn\cite{fb-unicorn} are two horizontally scalable systems which
are deployed at large scale in production.  TAO's query model  is
much more restricted than A1 in that it's not meant for large multi-hop
queries and it doesn't offer any consistency or atomicity guarantees.  Unicorn is built more  as a
search engine with very limited OLTP capability, but highly efficient
exploration queries like A1.  Since TAO and Unicorn are disk based,
their storage capacity is much larger than A1's.

LinkedIn's economic graph database \cite{linkedin-liquid} is a very
high performance graph query system designed for low latency queries
similar to A1. It scales up vertically and can answer lookup queries
in nanoseconds while A1 operates in microseconds.  Overall, A1 has
taken the approach of using cheap commodity hardware to scale out
while taking advantage of RDMA to keep query latency low, while the
LinkedIn database relies upon fixed sharding and specialized hardware
to achieve its performance.

As the price of RAM has fallen, building distributed in-memory storage
systems\cite{ramcloud} for low-latency applications has become very
attractive. The combination of RAM storage and RDMA networks is a
newer development and research systems like
FaRM\cite{farm-nsdi,farm-sosp} and NAMDb\cite{kraska17,kraska19} have shown the
advantages of using RDMA to build scale-out transactional
databases.  NAMDb has studied in detail the performance benefits of building remote pointer based data structures like B+ trees over RDMA.  The challenges of designing data structures optimized for remote memory has been considered by Aguiler et.al. \cite{far-memory,remote-memory}.  RDMA has been used to build high performance RDF
engines\cite{wukong} and file systems as well\cite{polarfs}. But the
adoption of RDMA in industry has been limited by the fact that it is
hard to ensure proper fairness and congestion control  in large data
center deployments with commodity hardware\cite{dcqcn}.  We believe the example
of A1 will augment the case for wide-spread adoption of RDMA in
cloud data centers.

\section{Conclusion}\label{sec:conclusion}
Building a generic database is a complex problem. A1 was designed to work in a space with huge 
data volume, wide variety of data sources and update frequencies and strict requirements to 
perform queries with very low latency.

Distributed systems are complex to program and operate, and we chose to implement transaction
support to hide the complexities of availability, replication and durability in the face of machine failure. The connected nature of 
graph data made it even more important to ensure correctness at any time. In our experience, the developer
productivity was high due to the support of transactions. Furthermore, the natural property graph
model was intuitive and powerful to use for building search-oriented applications in Bing.

FaRM and A1 utilizes the benefits of RDMA to a great extent, and the performance achieved makes more
complex question answering possible at scale, and within latencies acceptable for interactive searching. 
FaRM was originally designed to support relational systems, but our work also shows that it is general
enough to be considered a very efficient programming model for low-latency systems at scale.

\section{Acknowledgements}
A1 is built on top of the work by the FaRM team: in particular we'd like to thank Aleksandar Dragojevic, Dushyanth Narayanan, Ed Nightingale and our product manager Dana Cozmei. 
Getting A1 into production would not have been possible without the help and support of the ObjectStore team: Sam Bayless, Jason Li,  Maya Mosyak, Vikas Sabharwal, Junhua Wang and Bill Xu. The feedback we received from our customers in Bing was invaluable in guiding our roadmap and features. We would also like to thank our anonymous reviewers whose 
feedback improved the paper in multiple ways.
\bibliographystyle{ACM-Reference-Format}
\bibliography{biblio}
\end{document}